\documentclass[aps,prd,amsmath,amssymb,12pt]{revtex4}
\usepackage{color}
\usepackage{graphicx}
\usepackage{bm}
\def\journal#1#2#3#4{{#1} {\bf #2}, #3 (#4)}
\newcommand{\be}{\begin{equation}}
\newcommand{\ee}{\end{equation}}
\newcommand{\bea}{\begin{eqnarray}}
\newcommand{\eea}{\end{eqnarray}}
\newcommand{\hf}{\frac12}
\newcommand{\nn}{\nonumber\\}
\def\eq#1{(\ref{#1})}
\def\la{\langle}
\def\ra{\rangle}

\def\Tr{{\mathrm{Tr}}}
\def\ord#1{{\cal O}\left(#1\right)}
\def\mr#1{{\mathrm{#1}}}
\def\v#1{{\bm{#1}}}
\def\dt{\Delta t}
\def\dv#1{\dot{\v{#1}}}

\def\hpsi{{\hat\psi}}
\def\hpsid{{\hat\psi^\dagger}}

\def\hv#1{\hat{\v{#1}}}
\def\hx{\hat x}
\def\hy{\hat y}
\def\hD{{\hat D}}
\def\hG{{\hat G}}
\def\ih{\frac{i}{\hbar}}

\begin{document}
\title{Time scale of stationary decoherence}
\author{Janos Polonyi}
\email{polonyi@iphc.cnrs.fr}
\affiliation{Strasbourg University, CNRS-IPHC, 23 rue du Loess, BP28 67037 Strasbourg Cedex 2, France}

\begin{abstract}
The decoherence of a test particle interacting with an ideal gas is studied by the help of the effective Lagrangian, derived in the leading order of the perturbation expansion and in order $\ord{\partial^2_t}$. The stationary decoherence time is found to be comparable to or longer than the diffusion time. The decoherence time reaches its minimal value for classical, completely decohered environment, suggesting that physical decoherence is slowed down as compared with diffusion by the quantum coherence of the environment.
\end{abstract}
\maketitle

\section{Introduction}
Decoherence, which is the suppression of the interference between certain components of a quantum state \cite{zehd,zurekd}, plays an important role in the quantum-classical transition \cite{joos,zurekt}, and quite generally  its proper understanding poses a challenge. The problem is the pin down of the similarities and the differences between decoherence and diffusive processes. While both lead to a loss of information, realized by the increasing irrelevance in time of the initial conditions, they differ substantially; the decoherence being more indirect in its appearance in observables. A clear sign of their difference is their timescale, the decoherence is supposed to be several orders of magnitude faster than dissipation \cite{joos}. The importance of this result is that it suggests that the decoherence is quickly completed as a system reaches the classical scales and the ensuing relaxation can be considered as a classical process. In other words, there are two kinds of dissipation, a quantum and a classical, with two different mechanisms.

The decoherence builds up in time and its description is based on our understanding of open quantum systems. The simplest way open quantum systems can be handled is to assume some dissipative terms in the equation of motion for the density matrix. A local equation in time, called the master equation, was developed for a generic harmonic oscillator \cite{dekker,li} and was also used in quantum optics \cite{gardiner,savage,agarwal,agarwalk,dattagupta,sandulescuk,isarss,isark}. A more realistic  but still exactly soluble model consists of a system of linearly coupled harmonic oscillators \cite{agarwal,caldeira,unruh,grabert} where a non-Markovian memory term can be found \cite{hu92,hu93,hu94}, too. The master equation of a test particle, interacting with a gas has been derived by treating the particle-gas interaction as a sequence of collisions and assuming a stationary off-diagonality in the seminal work \cite{joos}, followed by the description of dissipation \cite{gallis} and the simultaneous inclusion of the decoherence and the friction forces in the equation of motion \cite{diosi,altenmuller,hornbergers,adler}. A microscopic introduction of dissipation was aimed by the help of the quantum linear Boltzmann equation with cross sections evaluated in the Born approximation \cite{vacchinie,lanz,vacchini,vacchinij} or without assuming the perturbation expansion \cite{hornbergerk}, cf. ref. \cite{vacchinis} for a summary of this scheme. The results of this treatment are in agreement with the master equation obtained within the framework of the traditional perturbative many-body technique \cite{dodd}. It has been pointed out that the decoherence can be characterized in different manners \cite{dyndec}. The stationary decoherence scheme, which consists of ignoring the internal system dynamics, has usually been considered \cite{joos} and is discussed here, too. While this approximation is acceptable for translation invariant systems it fails badly in the presence of harmonic forces \cite{dyndec}, e.g., a test particle, bound by a harmonic potential. The decoherence of a test particle, interacting with a gas, has been studied experimentally and a nice  demonstration of the loss of coherence of fullerenes \cite{hornbergerin} indicates that each collision can lead to an almost complete loss of coherence and lends a support to the collisional approach in describing decoherence.

Our goal in this work is a more systematic and careful way of estimating the stationary decoherence time. This is achieved within the closed time path CTP formalism which is well suited to deal with open systems. Its distinguishing feature; namely, the rather unusual reduplication of the degrees of freedom, is actually an efficient method of representing the environment of an open system. In fact, the environment is usually much larger and more complex than the observed system, hence the compression of the environment into a CTP copy of the system without a loss of relevant information is a highly nontrivial achievement. Our approach is based on the effective Lagrangian of a test particle, interacting with an ideal gas, which has already been calculated within the CTP formalism \cite{schw,keldysh,kamenev,rammer,calzetta} in a systematic approximation scheme, in the leading, $\ord{g^2}$, $\ord{x^2}$, $\ord{\partial_t^2}$ order of the perturbation expansion in the particle-gas interaction, characterized by the coupling strength, $g$, and the Landau-Ginzburg double expansion \cite{gas}. The test particle-gas entanglement, appearing in the order $\ord{g^4}$ is therefore ignored. It is found that the fast decoherence rate is predicted only if the perturbative expression is used beyond its domain of applicability. The more careful treatment of the approximate equations predicts that this decoherence timescale is equal to or shorter than the dissipative timescale. 

The earlier predictions about the faster decoherence can be traced back to the use of expressions beyond their limit of validity, namely the application of microscopic equations with macroscopic parameters. The size of a fullerene molecule, used as the test particle in the experimental verification of decoherence \cite{hornbergerin}, is microscopic, hence the predictions of the collisional approach to the macroscopic regime can not be tested by this method. It remains an  interesting and challenging task to follow the decoherence as the size of the test particle reaches the macroscopic regime and to improve the experimental method until it resolves the time evolution of the build up of the decoherence.

The presentation starts in section \ref{statdecs} with the introduction of the Green's function for the density matrix and the stationary decoherence approximation and continues with the derivation of the effective Lagrangian of the test particle in an ideal fermi gas and for a photon environment in sections \ref{idgasenvs} and \ref{photons}, respectively. The comparison of the conditions, needed to be satisfied in the collisional approach and in the calculation of the effective theory are surveyed in section \ref{compars}. The summary is given in section \ref{concls}. Three appendixes are added with a succinct review of the collisional approach to stationary decoherence, with the technical details of the derivation of the effective Lagrangian of the test particle in an ideal-gas environment and with the derivation of the master equation.

\section{Liouville space propagator and the decoherence}\label{statdecs}
The reduced density matrix of a degree of freedom, described by the coordinate $x$,
\be\label{effgen}
\rho_{t_f}(x^+,x^-)=\sum_n\la x^+|\la n|e^{-\ih H(t_f-t_i)}\rho_{t_i}e^{\ih H(t_f-t_i)}|x^-\ra|n\ra,
\ee
where the sum is over an environment basis, can formally be written in the form
\be
\rho_{t_f}(x^+_f,x^-_f)={\cal U}(t_f-t_i)\rho_{t_i}(x^+_i,x^-_i).
\ee
with ${\cal U}$ denoting the propagator in the Liouville space. The matrix elements of $\cal U$ are given in terms of the path integral, 
\be\label{lsprop}
\la\hx_f|{\cal U}(t)|\hx_i\ra=\int D[\hx]e^{\ih S_{eff}[\hx]},
\ee
where the integration extends over pairs of trajectories, $\hx=(x^+,x^-)$, where $x^+(t)$ and $x^-(t)$ is used in the path integral for the time evolution operator, $\exp-iHt/\hbar$ and its Hermitian conjugate, respectively, with fixed endpoints, $\hx(t_1)=\hx_i$, $\hx(t_2)=\hx_f$. The dressing of the effective action, $S_{infl}=S_{eff}-S_0$, the  influence functional, is complex for open systems and the decay of the Liouville space propagator, generated by $\mr{Im}S_{eff}$, is the manifestation of decoherence in the coordinate representation. The parametrization, $x^\pm=x\pm x^d/2$, where $x$ denotes the physical coordinate and $x^d$ stands for the quantum fluctuations, will be used frequently.

The decoherence is generated by $\mr{Im}S_{infl}>0$, evaluated for trajectories with large $|x^d(t)|$ and its description requires the solution of the full dynamical problem of the observed system and its environment. The decoherence arises from the orthogonalization of two relative environment states \cite{everett}, belonging to two system states, hence a natural approximation to this involved problem is the stationary decoherence scenario where the system dynamics is ignored. This is usually realized by solving a simplified master equation for the reduced density matrix \cite{joos},
\be\label{zehsme}
\dot\rho(x^+,x^-;t)=-F(x^+-x^-)\rho(x^+,x^-;t).
\ee
The same decoherence scheme is realized within the effective theory by approximating the path integral \eq{lsprop} by the value of its integrand, evaluated for a static trajectory pair, $x^\pm(t)=x^\pm$.

\section{Ideal gas environment}\label{idgasenvs}
The CTP formalism has already been used to derive the effective Lagrangian of a particle, moving in an ideal Fermi gas, in the leading order of the perturbation and the Landau-Ginzburg double expansion \cite{gas}. The action of a test particle, interacting with an ideal gas is written as a sum, $S=S_p+S_g$, with \bea\label{gasact}
S_p[\hv{x}]&=&\sum_\sigma\sigma\int dt\left[\frac{m_B}2\dot{\v{x}}^{\sigma2}(t)-U(\v{x}^\sigma(t))\right],\nn
S_g[\hv{x},\hpsid,\hpsi]&=&\sum_{\sigma\sigma'}\int dtd^3xdt'd^3x'\psi^{\sigma\dagger}(t,\v{x})(F^{-1})^{\sigma\sigma'}(t-t',\v{x}-\v{x}')\psi^{\sigma'}(t',\v{x}')\nn
&&-\sum_\sigma\sigma\int dtd^3x\psi^{\sigma\dagger}(t,\v{x})\psi^\sigma(t,\v{x})V(\v{x}-\v{x}^\sigma(t)),
\eea
where $\psi(t,\v{x})$ denotes the field operator controlling the gas particles in second quantization and $\sigma,\sigma'=\pm$. The simplectic structure, the sign difference between the $+$ and the $-$ contributions to the action is due to the opposite signs in the exponents in eq. \eq{effgen}. The propagator of the gas particles,
\bea\label{gasprop}
\hat F_{\omega,\v{k}}&=&\int d^4xe^{-i\omega x^0+i\v{k}\v{x}}F(x)\nn 
&=&\begin{pmatrix}\frac1{\omega-\epsilon_\v{k}+i\epsilon}-i2\pi\delta(\omega-\epsilon_\v{k})\xi n_\v{k}&-i\xi2\pi\delta(\omega-\epsilon_\v{k})n_\v{k}\cr-i2\pi\delta(\omega-\epsilon_\v{k})(1+\xi n_\v{k})&\frac1{\epsilon_\v{k}-\omega+i\epsilon}-i2\pi\delta(\omega-\epsilon_\v{k})\xi n_\v{k}\end{pmatrix},
\eea
contains the one single energy, $\epsilon_\v{k}=\hbar^2\v{k}^2/2m-\mu$, and the occupation number, $n_\v{k}=1/(e^{\beta(\epsilon_\v{k}-\mu)}-\zeta)$, for bosons ($\zeta=1$) or fermions ($\zeta=-1$). 

The reduced density matrix of the test particle, \eq{effgen} with $H$ being the Hamiltonian of the action \eq{gasact}, can easily be obtained in the path integral representation. The usual slicing of time, $t\to t+\dt$, applied for both the time evolution operator on the right hand side of eq. \eq{effgen}, produces
\be\label{genfgase}
\rho_{t_f}(\v{x}^+_f,\v{x}_f^-)=\int D[\hv{x}]e^{\ih S_p[\v{x}^+]-\ih S_p[\v{x}^-]+\ih S_{infl}[\hv{x}]},
\ee
where the convolution with the initial density matrix at $t_i=-\infty$ is suppressed. The integration is over particle trajectories, $\hv{x}(t_f)=\hv{x}_f$, ending at the desired matrix elements of the reduced density matrix and the influence functional, $S_{infl}[\hv{x}]$, is defined by integrating over field configurations which are made closed by the trace operation on the environment at $t'_f\ge t_f$, $\psi^+(t'_f,\v{x})=\psi^-(t'_f,\v{x})$, $\psi^{\dagger+}(t'_f,\v{x})=\psi^{\dagger-}(t'_f,\v{x})$, in
\be\label{inflfuncpg}
e^{\ih S_{infl}[\hv{x}]}=\int D[\hpsi]D[\hpsid]e^{\ih S_g[\hv{x},\hpsid,\hpsi]}.
\ee
The propagator, \eq{gasprop}, corresponds to the limit $t'_f\to\infty$. We go beyond the leading-order approximation of ref. \cite{gas} to capture the full dependence of the effective Lagrangian on the instantaneous off-diagonality.

\subsection{Influence functional}
The Gaussian integral,\eq{inflfuncpg}, is easy to carry out and its $\ord{V^2}$, leading order expression,
\be\label{inflg}
S_{infl}[\hv{x}]=-\hf\sum_{\sigma\sigma'}\sigma\sigma'\int dtdt'\Gamma^{\sigma\sigma'}\left(t',\v{x}^\sigma\left(t+\frac{t'}2\right)-\v{x}^{\sigma'}\left(t-\frac{t'}2\right)\right),
\ee
is given in terms of the bi-local Lagrangian,
\be\label{gamma}
\hat\Gamma(t,\v{x}-\v{x}')=\int d^3yd^3y'V(\v{x}-\v{y})\hG(t,\v{y}-\v{y}')V(\v{x}'-\v{y}')
\ee
where $G^{\sigma_1\sigma_2}(x_1-x_2)=-i\hbar\hat F^{\sigma_1\sigma_2}(x_1-x_2)\hat F^{\sigma_2\sigma_1}(x_2-x_1)$ denotes the particle-hole propagator. 

The functions $\hG$ and $\hat\Gamma$ of the influence functional display the block structure of a CTP two point function,
\be\label{blockg}
\hG=\begin{pmatrix}G^n+iG^i&-G^f+iG^i\cr G^f+iG^i&-G^n+iG^i\end{pmatrix},
\ee
containing the near and the far Green's functions, $G^n$ and $G^f$, as well as the imaginary part, $G^i$. We shall need the components $G^f_q=(G^{-+}_q-G^{-+}_{-q})/2$ and $iG^i_q=(G^{-+}_q+G^{-+}_{-q})/2$ of the particle-hole two-point function, given by 
\be
G^{-+}_{\omega,\v{q}}=-i\frac{2n_s}\hbar\int\frac{d^3q}{(2\pi)^3}2\pi\delta\left(\omega-\frac{\hbar\v{q}^2}{2m}+\frac{\hbar\v{k}\v{q}}m\right)n_\v{k}(1-n_{\v{k}-\v{q}}),
\ee
for an ideal Fermi gas, where $n_s$ stands for the spin degeneracy. Since $G^{-+}_{\omega,\v{q}}$ is not analytic at vanishing temperature we consider the gas at finite temperature where,
\be\label{gmp}
G^{-+}_{\omega,\v{q}}=\frac{n_sk_BTm^2}{2\pi\hbar^4|\v{q}|}\int_0^\infty dz\frac1{ae^z+be^{-z}+c}
\ee
with
\be
a=e^{\frac1{k_BT}(\frac{m\omega^2}{2\v{q}^2}+\frac{\hbar^2\v{q}^2}{8m}-\frac{\hbar\omega}2-\mu)},~~~
b=\frac{e^{-\beta\hbar\omega}}a,~~~
c=e^{-\beta\hbar\omega}+1.
\ee

The translation invariance of the environment restricts the dependence of the influence functional on the trajectory $\v{x}(t)$ to $\dv{x}(t)$ and higher order time derivatives. The $\ord{\partial_t^2}$ evaluation of the influence functional \eq{inflg} leads to the influence Lagrangian,
\be\label{anscont}
L_{infl}=\dv{x}\Delta m(\v{x}^d)\dv{x}^d-\dv{x}k(\v{x}^d)\v{x}^d+i\left[U_d(\v{x}^d)+\hf\dv{x}^dq(\v{x}^d)\dv{x}^d+\hf\dv{x}r(\v{x}^d)\dv{x}\right],
\ee
where $\Delta m$, $k$, $q$ and $r$ are $3\times3$ matrices. The $\v{x}^d$-dependence describes the environment induced modulation of the mass ($\Delta m$), friction constant ($k$), decoherence strengths ($q$, $r$) and the decoherence potential, $U_d$, controlling the stationary decoherence. When the time evolution of the reduced density matrix is considered  beyond the saddle point expansion, in the presence of quantum fluctuations, one encounters the problem of operator mixing. This ambiguity can be resolved by matching the influence action, 
\be
S^{reg}_{infl}=\dt\sum_nL^{latt}_{infl}(\hv{x}_{n+1},\hv{x}_n),
\ee
where $\hv{x}_n=\hv{x}(t_i+n\dt)$, and
\bea\label{dinfl}
L^{latt}_{infl}(\hv{x}_{n+1},\hv{x}_n)&=&\frac{\v{x}_{n+1}-\v{x}_n}\dt\Delta m\frac{\v{x}^d_{n+1}-\v{x}^d_n}\dt-\frac{\v{x}_{n+1}-\v{x}_n}\dt k\v{x}^d_{n+\xi}\nn
&&+i\left[U_d(\v{x}^d_{n+1})+\hf\frac{\v{x}^d_{n+1}-\v{x}^d_n}\dt q\frac{\v{x}^d_{n+1}-\v{x}^d_n}\dt+\hf\frac{\v{x}_{n+1}-\v{x}_n}\dt r\frac{\v{x}_{n+1}-\v{x}_n}\dt\right],
\eea
to eq. \eq{inflg} for small but finite $\dt$. All the $3\times3$ matrix functions $\Delta m$, $k$, $q$ and $r$ of the Lagrangian are evaluated at the intermediate point, $\hv{x}_{n+\xi}=(\hv{x}_{n+1}+\hv{x}_n)/2+\xi(\hv{x}_{n+1}-\hv{x}_n)$, $\xi$ being an additional, dimensionless parameter of the regularization. The functions in the effective Lagrangian are independent of $\v{x}$ owing to the translation invariance of the environment and the matching, outlined in appendix \ref{fullxds}, results in eqs. \eq{matching} for fermionic environment, $\zeta=-1$. To minimize the nonlocal effects, generated by the instantaneous potential, $V(\v{x})$, we consider contact interaction between the test particle and the gas, $V(\v{x})=g\delta(\v{x})$. 

In the case of strongly decohered motion the influence Lagrangian simplifies to the $\ord{\v{x}^{d2}}$, isotropic form \cite{gas},
\be\label{linflk}
L_{infl}=\Delta m\dv{x}^d\dv{x}-k\v{x}^d\dv{x}+\frac{i}2(d_0\v{x}^{d2}+d_2\dv{x}^{d2}),
\ee
with
\bea\label{qcouipl}
\Delta m&=&\frac1{12\pi^2}\int_0^\infty dqq^4|V_q|^2\partial_{i\omega}^2G^n_{0q},\nn
k&=&-\frac1{6\pi^2}\int_0^\infty dqq^4|V_q|^2\partial_{i\omega}G^f_{0q},\nn
d_0&=&-\frac1{6\pi^2}\int_0^\infty dqq^4|V_q|^2G^i_{0q},\nn
d_2&=&\frac1{12\pi^2}\int_0^\infty dqq^4|V_q|^2\partial_{i\omega}^2G^i_{0q},
\eea
involving a mass renormalization, $m=m_B+\Delta m$, and Newton's friction constant $k$. The constants $d_0$ and $d_2$ control the coordinate and the velocity dependent part the decoherence, respectively.

\subsection{Intrinsic decoherence scales}
We start with the dissipative timescale,
\be\label{dists}
\tau_{diss}=\frac{m}k,
\ee
to be found by using eq. \eq{gmp} for the calculation of the derivative,
\be
i\partial_\omega G^f_{\omega\v{q}|\omega=0}=\frac{n_sm^2}{2\pi\hbar^3|\v{q}|}\frac1{1+e^{\beta(\frac{\hbar^2\v{q}^2}{8m}-\mu)}},
\ee
to be used in the second equation of \eq{qcouipl} to find
\be\label{frtsc}
\frac1{\tau_{diss}}=\frac{32n_sg^2m}{3\pi\hbar^3\lambda^4_T}\int_0^\infty\frac{dzz}{1+e^{z-\frac{\epsilon_F}{k_BT}}},
\ee
where $\epsilon_F$ denotes the Fermi energy and $\lambda_T=\hbar\sqrt{2\pi/mk_BT}$ stands for the thermal wavelength.

The stationary decoherence approximation to eq. \eq{lsprop} consists of replacing the path integral by the value of the integrand at the static trajectory pair, $\hv{x}(t)=\hv{x}$, yielding the stationary decoherence suppression factor, 
\be
\left|e^{-\ih S_{infl}[\hv{x}]}\right|=e^{-\frac{T}{\tau_{sd}(\v{x}^d)}},
\ee
$T$ being the total time span of the propagation. The timescale, appearing in this expression when the influence Lagrangian \eq{anscont} is used,
\be
\tau_{sd}(\v{x}^d)=\frac\hbar{U_d(\v{x}^d)},
\ee
is not universal, being dependent on the off-diagonality $|\v{x}^d|$, and the form \eq{decpot} of the decoherence potential indicates that the decoherence length scale is the thermal wavelength, $\ell_{sd}=\lambda_T$.  It is instructive to compare the expression
\be
U_d(\v{x}^d)=\frac1{2\pi^2}\int_0^\infty dqq^2\left(\frac{\sin|\v{x}^d|q}{|\v{x}^d|q}-1\right)\Gamma^i_{0q}
\ee
of the decoherence potential with eq. \eq{scattfu}. Their similarity shows that the derivation of the effective theory and the collisional approach run parallel. Furthermore, the decoherence potential generates the right-hand side of the master equation, \eq{zehsme}, with  $F(\v{x}^d)=U_d(\v{x}^d)/\hbar$. 

\begin{figure}
\includegraphics[scale=0.8]{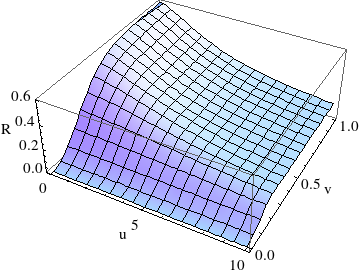}
\caption{The ratio $\tau_{diss}/\tau_{sd}$ [eq. \eq{dissdecrat}] plotted on the plane $(u,v)=(\epsilon_F/k_BT,x^d/\lambda_T)$.}\label{rf}
\end{figure}

The form \eq{gmp} of the off-diagonal CTP block of the particle-hole propagator can be used to calculate the static decoherence timescale in terms of the microscopic parameters,
\be\label{dectsig}
\frac1{\tau_{sd}(\v{x}^d)}=\frac{8n_sg^2m}{\pi\hbar^3\lambda_T^4}\int_0^\infty dz\frac{1-\frac{\sin4\sqrt{\pi z}\frac{|\v{x}^d|}{\lambda_T}}{4\sqrt{\pi z}\frac{|\v{x}^d|}{\lambda_T}}}{e^{z-\frac{\epsilon_F}{k_BT}}+1}
\ee
which together with eq. \eq{frtsc} yields the timescale ratio,
\be\label{dissdecrat}
\frac{\tau_{diss}}{\tau_{sd}(\v{x}^d)}=R\left(\frac{\epsilon_F}{k_BT},\frac{|\v{x}^d|}{\lambda_T}\right),
\ee
given by the dimensionless function
\be
R(u,v)=\frac34\frac{\int_0^\infty dz\frac{1-\frac{\sin4\sqrt{\pi z}v}{4\sqrt{\pi z}v}}{1+e^{z-u}}}{\int_0^\infty dz\frac{z}{1+e^{z-u}}}.
\ee
The state of the environment is characterized by two parameters, the temperature and the density. The variable $u=\epsilon_F/k_BT=\hbar^2k^2_F/2mk_BT$ is a dimensionless measure of the quantum nature of the environment, the ideal gas is in a pure state for $u=\infty$ and realizes a completely decohered, classical Gibbs ensemble when $u=0$. The variable $v=|\v{x}^d|/\lambda_T$ is the off-diagonality, expressed in the natural length scale of the environment.

\begin{figure}
\includegraphics[scale=0.6]{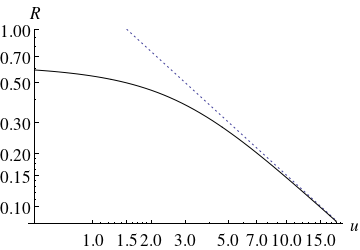}
\caption{The ratio $\tau_{diss}/\tau_{sd}$ [eq. \eq{dissdecrat}] shown against $u=\epsilon_F/k_BT$ for $x^d\gg\lambda_T$  (solid line). The dotted line corresponds to $3k_BT/2\epsilon_F$.}\label{af}
\end{figure}

The ratio, plotted in Fig. \ref{rf}, is a monotonic function of both $u$ and $v$, the stationary decoherence speeds up with respect to the dissipation with increased off-diagonality or more classical environment. The dependence on the off-diagonality defines two different regimes, $R(u,v)\sim2\pi v^2+\ord{v^3}$ for small $v$ in the weak off-diagonal regime, $|\v{x}^d|\ll\lambda_T$, and the limit $v\to\infty$ gives $R(u,v)=9\ln2/\pi^2+\ord{u}$ as $u\to0$ and $R(u,v)=3/2u$ for $u\to\infty$ in the strong off-diagonality regime, $|\v{x}^d|\gg\lambda_T$. The ratio \eq{dissdecrat}, 
\be
\frac{\tau_{diss}}{\tau_{sd}(\v{x}^d)}=2\pi\frac{\v{x}^{d2}}{\lambda_T^2}=\frac{mk_BT}{\hbar^2}\v{x}^{d2}.
\ee
is $\epsilon_F$-independent for weak off-diagonality. The stationary decoherence speeds up with increased off-diagonality and reaches an $\v{x}^d$-independent strength at strong off-diagonality, in agreement with the results found by the collisional method \cite{joos}. The saturated value of the ratio, 
\be\label{strodd}
\frac{\tau_{diss}}{\tau_{sd}}=\begin{cases}\frac{9\ln2}{\pi^2}\left[1-\ord{\frac{\epsilon_F}{k_BT}}\right]&k_BT\gg\epsilon_F,\cr\frac{3k_BT}{2\epsilon_F}=\frac{3mk_BT}{\hbar^2k_F^2}&k_BT\ll\epsilon_F,\end{cases}
\ee
with $k_F=\sqrt{2m\epsilon_F}/\hbar$, cf. fig. \ref{af} (b), indicates a maximal stationary decoherence strength where the ratio $\tau_{diss}/\tau_{sd}$ assumes a universal value. The independence of the saturated value from physical constants suggests a common origin of dissipation and decoherence, realized in its  maximal strength. The difference between the dissipative and the decoherence timescales appears when the environment regains some quantum features. The slowing down of decoherence with respect to diffusion is in agreement with the absence of the $\ord{\v{x}^{d2}}$ decoherence at vanishing temperature \cite{gas} and is natural for static trajectories in an environment which is in its ground state. This circumstance makes it plausible that the characteristic length scale, separating the $\ord{\v{x}^{d2}}$ and the saturated regimes, is $\lambda^{-1}_T$ rather than the other environment length scale, $k^{-1}_F$.

\subsection{Build up time}
The stationary decoherence belongs to a static $\v{x}^d$ trajectory. But there is a finite time between the initial conditions and the observation hence it is natural to inquire about the length of time needed by $\v{x}^d(t)$ so that the stationary decoherence approximation applies. Since the suppression is driven by the overlap of two relative states of the environment with separation $\v{x}^d$ an estimate is the time, $\tau_e(\v{x}^d)$, needed to build up two such states from a common initial one. This process is carried out by propagating excitations of the environment, described by the off-diagonal blocks of the environment CTP propagators. The genuine excitations, the quasiparticles, are absent in this case because they cannot be generated by the static chronon pair. What is left is to rely on the thermal excitations of the environment and one expects $\tau_{env}(\v{x}^d)\sim|\v{x}^d|/v_T$, where $v_T=\sqrt{k_BT/m}$ is the thermal velocity.

To check this scenario we introduce an IR cutoff, $f(t)=e^{-t^2/\tau_{IR}^2}$, which stops the build up of the environment states after a time $\tau_{IR}$. The decoherence potential, calculated with this cutoff, is
\bea
U_d(\v{x}^d,\tau_{IR})&=&-\int dt\int\frac{d\omega d^3q}{(2\pi)^4}f(t)\Gamma^i_{\omega\v{q}}e^{-i\omega t}(\sin\v{q}\v{x}^d-1)\nn
&=&-\frac{\tau_{IR}v^2_T}{4\pi^{5/2}}\int d\omega\int dqq^2e^{-\frac18\tau_{IR}^2\omega^2}G^i_{\omega q}\left(\frac{\sin qx^d}{qx^d}-1\right).
\eea
The use of eq. \eq{gmp} yields
\bea
U_d(\v{x}^d,\tau_{IR})&=&\frac{\tau_{IR}v_T^2k_BTm^2}{8\pi^{7/2}\hbar^4}\int_{-\infty}^\infty d\omega\int_0^\infty dqdz\frac{qe^{-\frac18\tau_{IR}^2\omega^2}(1-\frac{\sin qx^d}{qx^d})}{e^{z+w_-(\omega,q)}+e^{-z-w_+(\omega,q)}+e^{-\frac{\hbar\omega}{k_BT}}+1},
\eea
with 
\be
w_\pm(\omega,q)=\frac1{k_BT}\left(\frac{m\omega^2}{2q^2}+\frac{\hbar^2q^2}{8m}\pm\frac{\hbar\omega}2-\mu\right).
\ee
This rather lengthy expression shows that the IR cutoff spreads the frequency around zero and recohers the modes $q<1/\tau_{IR}v_T$, i.e., at distances $|\v{x}^d|>\tau_{IR}v_T$. Hence we indeed need the time $\tau_{env}(\v{x}^d)\sim|\v{x}^d|/v_T$ to reach decoherence at separation $\v{x}^d$.

\section{Photon environment}\label{photons}
The environment, supporting particle modes with linear dispersion relation, generates singular effective dynamics, the short distance effective interactions are stronger than for an ideal gas and the fixed velocity of propagation makes the build up of the stationary decoherence depending stronger on the IR cutoff. We explore now these issues in the case of a point charge moving in a photon bath of temperature $T$. The action is chosen to be the sum $S=S_r+S_M$, containing the action of the free particle,
\be
S_r=-mc\int ds,
\ee
the Maxwell action in Feynman gauge and the minimal coupling,
\be
S_M=-\frac1{4c}\int dxF_{\mu\nu}(x)F^{\mu\nu}(x)-\frac1{2c}\int dx(\partial_\mu A^\mu(x))^2-\frac{e}c\int ds\dot x^\mu A_\mu(x(s)),
\ee
with $F_{\mu\nu}=\partial_\mu A_\nu-\partial_\nu A_\mu$. The photon propagator,
\be
\hD_q=\begin{pmatrix}\frac1{q^0+i\epsilon}&-2\pi i\delta(q^2)\Theta(-q^0)\cr
-2\pi i\delta(q^2)\Theta(q^0)&-\frac1{q^0-i\epsilon}\end{pmatrix}-i2\pi\delta(q^2)n_{|\v{q}|}\begin{pmatrix}1&1\cr1&1\end{pmatrix},
\ee
contains the Planck distribution, $n_q=2/(e^{\hbar cq/k_BT}-1)$. The cutoff-induced instabilities \cite{cutoff} can be ignored for stronger off-diagonality than the minimal distance, the UV cutoff, hence the naive, unregulated photon propagator can be used.

\begin{figure}
\includegraphics[scale=.8]{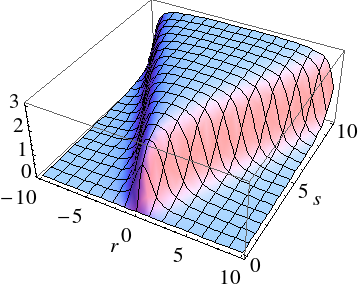}
\caption{The integrand of eq. \eq{imlinflphot}, plotted vs $(r,s)=(ct/\lambda_{T\gamma},|\v{x}^d|/\lambda_{T\gamma})$.}\label{photonf}
\end{figure}

The result of the Gaussian integral over the photon field can be found by eliminating the vector potential by the help of its equation of motion and one finds the influence functional
\be\label{inflactch}
S_{infl}[\hx]=\frac{e^2}{2c}\sum_{\sigma\sigma'}\sigma\sigma'\int dsds'\dot x^{\sigma\mu}(s)D^{\sigma\sigma'}(x^\sigma(s)-x^{\sigma'}(s'))\dot x^{\sigma'}_\mu(s'),
\ee
which can be transformed into the form \eq{inflg} with
\be
\Gamma^{\sigma\sigma'}(t,\v{x}-\v{x}')=-e^2cD^{\sigma\sigma'}\left(u,\v{x}^\sigma\left(t+\frac{u}2\right)-\v{x}^{\sigma'}\left(t-\frac{u}2\right)\right),
\ee
where $s=ct$, $x^\mu=(ct,\v{x})$ We assume a static world line pair, $x^\pm(s)=x\pm x^d/s$, and elementary steps lead to,
\be\label{gammaphot}
\Gamma^i(t,\v{x}^d)=\frac{e^2c}{4\pi}P\frac1{c^2t^2-\v{x}^{d2}}+\frac{e^2c}{4\pi^2\lambda_{T\gamma}|\v{x}^d|}\left[f\left(\frac{|\v{x}^d|-ct}{\lambda_{T\gamma}}\right)+f\left(\frac{|\v{x}^d|+ct}{\lambda_{T\gamma}}\right)\right],
\ee
where $\lambda_{T\gamma}=\hbar c/k_BT$ and
\be\label{fintdph}
f(y)=\int_0^\infty dz\frac{\sin zy}{e^{z}-1}\approx\arctan\left(y\frac{\pi^2}6\right),
\ee
the error of the approximation being a few percent for $y\sim1$ and vanishing as $y\to0$ or $\pm\infty$. The static charge decouples from the radiation field, reflected in the vanishing of the first term on the right-hand side of eq. \eq{gammaphot}, standing for the vacuum contribution to the influence functional. The rest gives
\be\label{imlinflphot}
U_d(\v{x}^d)=\frac{e^2c}{8\pi^2\lambda_{T\gamma}|\v{x}^d|}\int dt\left[f\left(\frac{|\v{x}^d|-ct}{\lambda_{T\gamma}}\right)+f\left(\frac{|\v{x}^d|+ct}{\lambda_{T\gamma}}\right)\right],
\ee
together with $\ell_{sd}=\lambda_{T\gamma}$. The integrand, plotted in Fig. \ref{photonf}, indicates that the dominant contribution comes from spacetime points with acausal separation. This result may seem surprising but one should bear in mind that we see here the suppression, given by the overlap of the bra and ket components of the photon state in the full density matrix, the relative states of the bra and the ket system components. The two photon states correspond to two static charges displaced from each other by $\v{x}^d$. The photons leave the charge system after a time $|\v{x}^d|/c$ and no further suppression takes place. The plateau of the integrand indicates that we have to sustain the static separation, $\v{x}^d$, for a time $|\v{x}^d|/c$ to recover the full decoherence strength which builds up linearly in time, in a manner similar to the case of the fermi gas environment. The approximated form of the integral \eq{fintdph} yields $U(\v{x}^d)\approx e^2/4\pi\lambda_{T\gamma}$ and an $\v{x}^d$-independent stationary decoherence timescale,
\be\label{photsdect}
\tau_{sd}=\frac{\lambda_{T\gamma}}{\alpha c}=\frac\hbar{\alpha k_BT}
\ee
with $\alpha=e^2/4\pi\hbar c$. The thermal length scale, $\lambda_T\approx0.2/T$cm, $T$ being given in Kelvin, gives $\tau_{sd}\approx0.76\times10^{-9}/T$sec. 

The real part of the influence Lagrangian, \eq{inflactch}, has already been derived in $\ord{\hv{x}^2}$ \cite{point}. The Newtonian, $\ord{\dv{x}}$, form of the radiative friction force is canceled by Lorentz invariance, assuming that the photons are in the ground state. The Abraham-Lorentz force arises as an anomaly of a one-loop diagram, representing the eliminated classical electromagnetic field, and provides the dissipative timescale, 
\be\label{lorfrtsc}
\tau_{diss}=\frac{2\alpha\hbar}{3mc^2}<\tau_{sd},
\ee
where the inequality holds well beyond the pair creation threshold, up to the temperature $T_{cr}=3mc^2/2\alpha^2k_B$. Note that the Abraham-Lorentz force, being the result of the nonuniform convergence of the loop-integral, is independent of temperature since the latter influences the photon propagator at finite distance scales only.

\section{Collisional approach and the limits of applicability}\label{compars}
We are now in the position to compare the collisional approach to decoherence, summarized in appendix  \ref{coldecs}, with our scheme and inspect the domain of applicability of the different descriptions. Our derivation of the stationary decoherence potential is qualitatively similar to the construction, leading to the collision driven master equation, the cross section taking the place of the coupling strength. The comparison of the two schemes is the easiest with the help of the master equation whose derivation within the effective dynamics is presented in appendix \ref{masters}. The master equation, generated by the influence Lagrangian \eq{linflk} can be brought into the form \cite{gas}
\bea\label{masteq}
\partial_t\rho&=&\frac1{i\hbar}\left[\frac{\v{p}^2}{2m}+\frac{k}{4m}\{\v{x},\v{p}\},\rho\right]\nn
&&-\frac{d_0+\frac{d_2k^2}{m^2}}{2\hbar}[\v{x},[\v{x},\rho]]-\frac{ik}{2m\hbar}[\v{x},\{\v{p},\rho\}]-\frac{d_2k}{m^2\hbar}[\v{x},[\v{p},\rho]]-\frac{d_2}{2m^2\hbar}[\v{p},[\v{p},\rho]]
\eea
while the collision based approach of (stationary) decoherence relies on the same master equation with $d_2=0$. Furthermore eqs. \eq{qcouipl} yield $\hbar d_0/2k=k_BT$, leading to the ratio
\be\label{quadrdecapr}
\frac{\tau_{diss}}{\tau_{sd}(\v{x}^d)}=4\pi\frac{\v{x}^{d2}}{\lambda_T^2},
\ee
which is the key result of the collisional based approach.

How can the agreement of the two schemes on eq. \eq{quadrdecapr} be reconsolidated with the sharply different order of magnitude estimates this relation provides for the ratio $\tau_{sd}/\tau_{diss}$? The different conclusions, extracted from the collisional approach and the present work actually come from the different use of qualitatively similar equations. An extremely short decoherence is reported in the collisional approach for a macroscopic test particle, a dust grain. A larger test particle interacts with more gas particles making the cross section and its counterpart of the present scheme, the coupling strength $g^2$ in \eq{dectsig}, larger. The questions one faces here are the applicability of (i) the perturbation expansion and (ii) the expansion in $\v{x}^d$. 

(i) One assumes that the perturbation expansion in the particle-gas interaction and the independent scattering approximation are valid. An estimate of the dimensionless small parameter of the expansion in $g$ is the ratio of the decoherence contribution to the Lagrangian, $\mr{Im}L_{infl}=\hbar/\tau_{sd}$ and the average kinetic energy, $k_BT$. Thus the condition for using the perturbation expansion is 
\be\label{pertappl}
\frac\hbar{k_BT}\sim\frac{10^{-11}}T<\tau_{sd},
\ee
$T$ and $\tau_{sd}$ being given in kelvin and sec, respectively, making the prediction of the influence Lagrangian, arising from the interaction with an ideal fermi gas, unreliable for $\tau_{sd}<10^{-13}$sec at room temperature. The perturbation expansion is applicable in the photon gas owing to $\alpha<1$, cf. eq. \eq{photsdect}.

The applicability of the single collision approach, outlined in Appendix \ref{coldecs}, relies on several conditions. (a) First, we assume that the change of the density matrix during a single collision,  \eq{etah}, is small compared with the density matrix itself, $F\dt<1$. The time between two consecutive collisions is at least $r_0/v_e$, $r_0$ and $v_e$ denoting the average separation and the velocity of the environment particles, and the applicability of the collisional approach is limited by the inequality
\be\label{mindects}
\frac{r_0}{v_e}<\tau_{sd}.
\ee
In the case of an ideal fermi gas we have $m/\hbar k_F^2<\tau_{sd}$. For the air at normal pressure and temperature one finds $r_0=3\times10^{-7}$cm, $v_e=\sqrt{2k_BT/m}\sim10^5$cm/s, and $\tau_{sd}>10^{-12}$s. Finally, $r_0\sim c\hbar/k_BT$, $v_e=c$ in a photon gas and the inequality \eq{pertappl} is recovered. Note that the bound on the decoherence time is always given by the natural microscopic timescale of the environment. (b) Multiple scatterings, implying the inequality $r_0>\ell_0=1/\sigma_{tot}n_g$, $\ell_0$ being the mean-free path of the gas, are ignored in the derivation. The minimal decoherence time, given by eq. \eq{mindect}, is just at the threshold of the multiple scattering regime, $\tau_{dmin}\sim\ell_0m/\hbar q_F$, and the master equation, derived in this scheme, is not reliable in the saturated regime, $|\v{x}^d|>r_0$. Although the derivation of the master equation by the help of the quantum linear Boltzmann equation can be extended by replacing the Born amplitudes by the exact transition amplitudes \cite{hornbergerk}, the approximation of ignoring the multiparticle collisions is kept in the construction. (c) Yet another assumption of the derivation is the applicability of the limit $\dt\to0$. Since $\dt>\dt_{min}$, cf. eq. \eq{mindt}, the master differential equation, \eq{zehsme}, can not resolve the time dependence below $\dt_{min}$ and the condition \eq{mindects} is found again.

(ii) Another question to settle is the identification of the length scale, $\ell_{tr}$, where the decoherence potential changes from an $\ord{\v{x}^{d2}}$ form to a saturated, separation independent constant. According to Fig. \ref{rf} $\ell_{tr}\sim\lambda_T/2$ at high temperature and decreases with the temperature to approximately $\ell_{tr}\sim\lambda_T/4$. The gas particles cease to orthogonalize themselves in the relative environment state at separation beyond $|\v{x}^d|\sim\lambda_{tr}$ which by the help of eq. \eq{quadrdecapr} excludes $\tau_{diss}\gg\tau_{dec}(\v{x}^d)$. The effective cross section, \eq{effcrsect}, is unknown and the phenomenological, collisional approach can not accurately identify $\ell_{tr}$. 

Finally, we mention two assumptions, common in both approaches. One is related to the treatment of a solid object as a structureless, point-like particle. The internal structure assumes its own dynamics within the time duration $a/v_{ph}$ where $a$ denotes the size of the object and $v_{ph}$ stands for the speed of the collective excitations, phonons, within the solid and the bound, 
\be
\frac{a}{v_{ph}}<\tau_{sd},
\ee
follows. Another assumption, made in the stationary decoherence scenario, is that the off-diagonality is held constant until the full stationary decoherence strength is build up. The decoherence is the result of real, physical processes, taking place in the environment and the stationary decoherence strength is reached at the off-diagonality $\v{x}^d$ during the time $|\v{x}^d|/v_e$. Hence the inequality,
\be\label{statdcon}
|\dv{x}^d|<v_e,
\ee
represents the condition of ignoring the dynamics, represented by the $\ord{\partial_t^2}$ terms of the effective Lagrangian, in building up the decoherence. Sine $\v{x}^d$ represents the quantum fluctuations of the test particle position the assessment of the validity requires us to consider the dynamics of the particle.

\section{Conclusions}\label{concls}
The stationary decoherence of the coordinate of a test particle injected into an ideal gas is investigated in this paper by deriving the effective Lagrangian in the leading order, using the test particle-gas-interaction, the amplitude and the frequency of the distortion of the particle trajectory as small parameters. 

The dynamics of the test particle has several timescales which may make up the decoherence time. There is a dissipative timescale and the internal dynamics of the test particles may possess further timescales. The multiple scatterings are ignored in the collision-based calculations, which places a lower bound on the decoherence time, given by the time between two consecutive collisions of test and gas particles; the dissipative timescale. The mixing of the timescale can be more clearly followed in the calculation of the effective Lagrangian. There is a proliferation of scales in weakly coupled theories, the different powers of the small parameter, the dimensionless strength of interaction, multiplying the scales of the free system produce new characteristic scales. However, such a phenomenon is not taking place in the leading order calculation of the decoherence timescale, followed here, both the dissipative and the stationary decoherence scales being proportional to the coupling constant. It is found that the stationary decoherence timescale can not be shorter than the dissipative time. It remains to see whether strong interactions can reverse this conclusion. The minimal time stationary decoherence time is an lower bound for the true, physical decoherence time, too. 

The dissipation can already be found in classical physics hence the ratio of the dissipative and the stationary decoherence time must be $\hbar$ dependent. Such a dependence motivates the use of a tripartite scheme where the gas, realizing the environment, is coupled to a heat bath whose role is to control the classicality of the gas, the primary environment of the test particle. Both the dissipation and the decoherence are generated by the environment so one expects an identical mechanism for both if the gas is classical. Support of this scenario is found by monitoring the dependence of the ratio of the dissipative and the stationary decoherence timescales as a function of gas temperature. This ratio is found to be universal, independent of the physical parameters at high temperature. In other words, the difference between dissipation and decoherence is due to the quantum coherence of the environment. The ratio decreases as the gas is cooled, indicating that the environment looses its efficiency to decohere the system when its own  coherence is increased.

\acknowledgments
I thank J\'anos Hajdu and Martin Jan\ss en for several useful discussions.

\appendix
\section{Collisional decoherence}\label{coldecs}
The collision based approach to decoherence started with the seminal paper \cite{joos}, followed by refs.  \cite{gallis,diosi,adler,vacchini,lanz,vacchinie,hornbergerk,dodd}. The goal of these works is the master equation describing the time dependence of the reduced density matrix of a test particle, the system, interacting with a gas, the environment. The results of this approach are briefly reviewed in this appendix to make the comparison with the CTP formalism easier. The reader who only wishes to follow the CTP effective theory approach to decoherence may skip this appendix.

\subsection{Master equation}
We restrict ourselves to the limit when the test particle is much heavier than the particles of the gas, making the recoil of the test particle during collisions negligible and the test particle is not entangled with the gas. (These restriction can be removed within the CTP effective action scheme by including higher loop graphs.) Such a simplification can be exploited more easily in the coordinate representation \cite{hornbergers} where a pure factorized state can be written in the product form $|\v{x}\ra\otimes|\psi\ra$, the second factor standing for the state of the gas. We want to find the result of a collision, the state $S|\v{x}\ra\otimes|\psi\ra$ where $S$ denotes the scattering matrix and $|\v{x}\ra=e^{-\ih \v{p}\v{x}}|\v{0}\ra$. By exploiting the translation invariance, $[\v{p}_p+\v{p}_g,S]=0$, we write $S|\v{x}\ra\otimes|\psi\ra=e^{-\ih(\v{p}_p+\v{p}_g)\v{x}}S|\v{0}\ra\otimes e^{\ih\v{p}_g\v{x}}|\psi\ra$. In the next step one introduces the scattering matrix for the gas particles, $S_0$, assuming a static, nonrecoiling test particle, located at the origin and finds $S|\v{x}\ra\otimes|\psi\ra=|\v{x}\ra\otimes e^{-\ih\v{p}_g\v{x}}S_0e^{\ih\v{p}_g\v{x}}|\psi\ra$. The final result is that a single collision induces the change, $\rho(\v{x}^+,\v{x}^-)\to\rho(\v{x}^+,\v{x}^-)\eta(\v{x}^+,\v{x}^-)$, of the reduced density matrix of the test particle where the multiplicative factor is 
\be\label{colleta}
\eta(\v{x}^+,\v{x}^-)=\Tr_e[e^{-\ih\v{p}_g\v{x}^+}S_0e^{\ih\v{p}_g\v{x}^+}\rho_ge^{-\ih\v{p}_g\v{x}^-}S^\dagger_0e^{\ih\v{p}_g\v{x}^-}],
\ee
where the trace is taken over the Hilbert space of the gas and $\rho_g$ is the initial density matrix of the gas.

It is advantageous to introduce a large but finite quantization box of volume $V$ with normalized one-particle excited states $|\v{q}\ra_V$ in the intermediate steps of the calculation, yielding
\be\label{etae}
\eta(\v{x}^+,\v{x}^-)=\frac{(2\pi\hbar)^3}V\sum_\v{q}\mu(\v{q})e^{\ih\v{q}(\v{x}^--\v{x}^+)}\la\v{q}|S_0e^{\ih\v{p}_g(\v{x}^+-\v{x}^-)}S^\dagger_0|\v{q}\ra_V,
\ee
where $\mu(\v{q})=\la\v{q}|\rho_g|\v{q}\ra_V$ stands for density of states in the rest frame of the gas. By introducing $T$, $S_0=\openone+iT_0$, and the unitarity of $S_0$, $T_0T_0^\dagger=i(T^\dagger_0-T_0)$ eq. \eq{etae} assumes the form
\be
\eta(\v{x}^+,\v{x}^-)=\frac{(2\pi\hbar)^3}V\sum_\v{q}\mu(\v{q})\left[1-\la\v{q}|T_0T_0^\dagger|\v{q}\ra_V+e^{\ih\v{q}(\v{x}^--\v{x}^+)}\la\v{q}|T_0e^{\ih\v{p}_g(\v{x}^+-\v{x}^-)}T^\dagger_0|\v{q}\ra_V\right]
\ee
which in turn can be written as
\be\label{colletacont}
\eta(\v{x}^+,\v{x}^-)=\frac{(2\pi\hbar)^3}V\sum_\v{q}\mu(\v{q})\left[1-\sum_{\v{q}'}\left(1-e^{\ih(\v{q}-\v{q}')(\v{x}^--\v{x}^+)}\right)|\la\v{q}|T_0|\v{q}'\ra_V|^2\right].
\ee
The continuum notation is at least partially restored in the form
\be
\eta(\v{x}^+,\v{x}^-)=1-\frac{(2\pi\hbar)^3}V\int d^3qd^3q'\mu(\v{q})\left(1-e^{\ih(\v{q}-\v{q}')(\v{x}^--\v{x}^+)}\right)|\la\v{q}|T_0|\v{q}'\ra|^2,
\ee
by using $\int d^3q\mu(\v{q})=1$. Therefore a single collision induces the change $\rho\to\rho+\Delta\rho$ of the reduced density matrix with
\be\label{etak}
\Delta\rho=-\rho(\v{x}^+,\v{x}^-)\frac{(2\pi\hbar)^3}V\int d^3qd^3q'\mu(\v{q})\left(1-e^{\ih(\v{q}-\v{q}')(\v{x}^--\v{x}^+)}\right)|\la\v{q}|T_0|\v{q}'\ra|^2.
\ee
In terms of the time evolution this implies $\rho(t+\Delta t)=\rho(t)+\Delta\rho$ for $\Delta t>r_0/v_e$ being the time between two consecutive collisions, expressed by the help of the typical  separation and the velocity of the particles in the gas, respectively.

The matrix element, appearing in \eq{etak} contains a distribution,
\be
\la\v{q}|T_0|\v{q}'\ra=\frac{i}{2\pi\hbar m}\delta(E(\v{q})-E(\v{q}'))f(\v{q},\v{q}')
\ee
and its square requires special care. The idea, leading to derive Fermi's Golden Rule can be used again to write one of the distributions as \cite{adler},
\be\label{fermi}
\delta(E)\approx\frac1{2\pi\hbar}\int_{-\frac{\Delta t}2}^{-\frac{\Delta t}2}dte^{\ih Et},
\ee
assuming $\Delta t>r_{sc}/v_e$, $r_{sc}$ being the distance scale of the interaction between the test and the gas particles. The resulting expression
\be
\delta^2(E)\approx\delta(E)\frac{\Delta t}{2\pi\hbar},
\ee
replaced into eq. \eq{etak} gives
\be\label{etah}
\Delta\rho(\v{x}^+,\v{x}^-)=-\rho(\v{x}^+,\v{x}^-)\frac{\Delta t}V\int d^3q\mu(\v{q})\frac{|\v{q}|}m\int d^2n\left(1-e^{\ih(\v{q}-\v{n}|\v{q}|)(\v{x}^--\v{x}^+)}\right)|f(\v{q},\v{n}|\v{q}|)|^2,
\ee
as the change of the density matrix due to a single collision where the second integral is over the unit sphere. The change after $N$ decohered collision is $N$ times of this expression,
\be\label{dmast}
\frac{\rho(\v{x}^+,\v{x}^-,t+\Delta t)-\rho(\v{x}^+,\v{x}^-,t)}{\Delta t}=-F(\v{x}^+-\v{x}^-)\rho(\v{x}^+,\v{x}^-),
\ee
with
\be\label{scattfu}
F(\v{x})=\int_0^\infty dq\nu(q)\frac{\hbar q}m\int\frac{d^2nd^2n'}{4\pi}\left(1-e^{\ih q(\v{n}-\v{n}')\v{x}}\right)|f(q\v{n},q\v{n}')|^2,
\ee
where $\nu(q)$ denotes the density of state in the absolute magnitude of the momentum. This latter is defined by the equation
\be
\mu(\v{q})d^3q=\frac1{4\pi}\frac{V}N\nu(q)dqd^2n,
\ee
and is normalized to the density of the gas, $\int dq\nu(q)=n_g$. The master equation \eq{zehsme} is obtained finally by taking the limit $\Delta t\to0$ in eq. \eq{dmast}. Note that this limit is symbolic only and the master equation, derived in this manner is lacking of the dynamics, taking place at times shorter than 
\be\label{mindt}
\Delta t_{min}=\max\left(\frac{r_0}{v_e},\frac{r_{sc}}{v_e}\right).
\ee
A further limitation on the time resolution is the assumption that all contributing individual scattering processes are completely decohered. However, the main problem with this master equation is the complete lack of the recoil, the dynamics of the test particle itself.

\subsection{Separation dependence}
The decoherence time, predicted by the master equation \eq{zehsme}, 
\be\label{cdects}
\tau_{sd}(\v{x}^+-\v{x}^-)=\frac1{F(\v{x}^+-\v{x}^-)},
\ee
depends on the off-diagonality in the coordinate representation and it is easy to separate two distinct regimes.

For small off-diagonality, $x^d=|\v{x}^+-\v{x}^-|\ll\hbar/p_g$, $p_g$ denoting the typical momentum scale of the one-particle excitations of the gas, one expands the exponential function in \eq{scattfu} and finds after the integration over the directions
\be
F(\v{x})=\v{x}^2\Lambda
\ee
where the coefficient
\be
\Lambda=\int_0^\infty dq\nu(q)\frac{\hbar q}m\frac{q^2}{\hbar^2}\sigma_{eff}(q)
\ee
is expressed with the help of an effective total cross section,
\be\label{effcrsect}
\sigma_{eff}(q)=\frac{2\pi}3\int d\cos\theta(1-\cos\theta)|f(q\v{z},q\v{n})|^2,
\ee
modulated by the factor $1-\cos\theta$ in the averaging over the scattering angle, $\theta$. The decoherence time in this regime,
\be
\tau_{sd}(x^d)=\frac1{x^{d2}\Lambda}.
\ee

The expression \eq{scattfu} is saturated for large separation and a lower bound for the decoherence time is provided by
\be
F=\int_0^\infty dq\nu(q)\frac{\hbar q}m\sigma_{tot}(q),
\ee
where
\be
\sigma_{tot}(q)=\int\frac{d^2nd^2n'}{4\pi}|f(q\v{n},q\v{n}')|^2,
\ee
denotes the total cross section at momentum $\hbar q$, averaged over the direction. Since the double integration of the unit sphere sums over the final directions and averages over the initial one reproducing the total cross section, $\sigma_{tot}(q)$, and the decoherence timescale is given by
\be\label{mindect}
\tau_{dmin}=\frac1{\int_0^\infty dq\nu(q)\frac{\hbar q}m\sigma_{tot}(q)}
\ee
and represents the saturated, minimal value of \eq{cdects}.

\subsection{Decoherence by photon scattering}
A simple model of a macroscopic object is a sphere of radius $a$ with dielectric constant $\epsilon$  \cite{joos}. According to the Rayleigh law 
\be
|f(q\v{z},q\v{n})|^2=q^4a^6\left(\frac{\epsilon-1}{\epsilon+2}\right)^2\frac{1+\cos^2\theta}2,
\ee
we have 
\be
\sigma_{eff}(q)=\frac{8\pi}9q^4a^6\left(\frac{\epsilon-1}{\epsilon+2}\right)^2
\ee
which together with the Planck distribution gives
\be
\Lambda=\frac{8\pi}98!\zeta(9)a^6c\left(\frac{\epsilon-1}{\epsilon+2}\right)^2\left(\frac{k_BT}{\hbar c}\right)^9,
\ee
where $\zeta(z)$ is Riemann's $\zeta$-function, $\zeta(9)=1.002$.

\section{Influence Lagrangian in the $\ord{\partial_t^2}$ derivative expansion}\label{fullxds}
The derivation of the influence Lagrangian \eq{anscont} from the translation invariant, non-local influence functional, \eq{inflg} is summarized in this appendix without expanding in the coordinate and keeping track of operator ordering ambiguities. The latter is absent and the continuous time formalism can safely be used in the path integrals for dynamics, generated by a Hamiltonian of the form $H=p^2/2m+U(x)$ only. The point here is that the operator mixing, the appearance of the products, $x^mp^n$, in the Hamiltonian, introduces UV divergences and the path integral formulas need a regulator; a small but finite time step, $\dt$. In fact, the Feynman propagator is $\ord{\omega^{-2}}$ for large frequency and generates a linear divergence for the velocity square, $\la\dot x^2\ra\sim\dt^{-1}$ \cite{rgqm}. Such a divergence leads to a dependence  of the expectation values, formed by the Lagrangian \eq{anscont}, on the way the functions $\Delta m$, $k$, $q$ and $r$ are defined at the scale of the cutoff, $\dt$. Therefore, the influence Lagrangian must be extracted for $\dt>0$. The influence Lagrangian, \eq{inflg}, was derived in the presence of an UV cutoff $E_{max}$, a maximal energy of the ideal gas dynamics. However, $E_{max}\dt\gg\hbar$ and its natural variable, the trajectory $\hv{x}(t)$, can be considered in continuous time as far as the low energy effective dynamics of the test particle is concerned. The influence Lagrangian will be found by matching it to the effective action, evaluated for the trajectory $\hv{x}(t)=\hv{x}+\hv{y}(t)$, where the fluctuation is orthogonal to the stationary part, 
\be\label{orth}
\int dt\hv{y}(t)=0.
\ee

The time derivative is replaced on a lattice by a finite difference operator, $\nabla_\pm f_n=\pm(f_{n\pm 1}-f_n)/\dt$ and a Lagrangian with time derivatives up to $\ord{\partial_t^M}$ describes correlations among $2M$ time slices. To handle such nonlocal terms one introduces a superlattice where $M$ consecutive sites of the original lattice are blocked into a single supersite, $\tilde n=\{M\tilde n,M\tilde n+1,\cdots,(M+1)\tilde n-1\}$, and a trajectory, $x_n$, develops $M$-components, corresponding to the first $M-1$ derivatives, $\tilde x_{\tilde n}=(x_{M\tilde n},\nabla_+x_{M\tilde n},\ldots,\nabla^{M-1}_+x_{M\tilde n})$, and the Lagrangian becomes first order in the time derivative when written in terms of the superlattice variables. Actually, we can continue to use the original lattice if the truncation is at $\ord{\partial_t^2}$ where our ansatz, \eq{anscont}, yields the nearest-neighbor interactions, given by the Lagrangian \eq{dinfl}.

To match the influence Lagrangian we start with a trajectory $\hv{x}_n$, used for \eq{dinfl}, and construct an interpolating trajectory, 
\be\label{intptr}
\hv{x}(n\dt+\tau)=\hv{x}_n+(\hv{x}_{n+1}-\hv{x}_n)\tau,
\ee
$0<\tau<1$, for the evaluation of the influence functional. For that end we need the Fourier transforms,
\bea
\tilde{\hv{x}}_\omega&=&\dt\sum_{n=-\infty}^\infty e^{i\omega\dt n}\hv{x}_n,\nn
\hv{x}_\omega&=&\int_{-\infty}^\infty dte^{i\omega t}\hv{x}(t),
\eea
related by the equation
\be
\v{x}_\omega=\tilde{\v{x}}_\omega[1+i\omega\dt+\ord{\dt^2}].
\ee

The matching of the stationary, $\ord{y^0}$, contributions results
\be\label{ordyzil}
U_d(\v{x}^d)=\int dt[\Gamma^i(t,\v{x}^d)-\Gamma^i(t,\v{0})]
\ee
and the particle-hole two-point function, \eq{gmp}, yields
\be\label{decpot}
U_d(\v{x}^d)=\frac{8m}{\pi\hbar^2\lambda_T^4}\int_0^\infty dz\frac{|V_{k_F\sqrt{z}}|^2}{e^{z-\frac\nu{4\pi}}+1}\left(1-\frac{\sin4\sqrt{\pi z}\frac{|\v{x}^d|}{\lambda_T}}{4\sqrt{\pi z}\frac{|\v{x}^d|}{\lambda_T}}\right),
\ee
where $\lambda_T=\sqrt{2\pi\hbar^2/mk_BT}$ is the thermal wavelength, the Fermi wave vector,  $k_F=\sqrt{2m\epsilon_F}/\hbar$, is given in terms of the Fermi energy, $E_F$, $\nu=\lambda_T^2k_F^2$, and 
\be
V_{|\v{q}|}=\int d^3xe^{-i\v{x}\v{q}}V(|\v{x}|)
\ee
denotes the Fourier transform of the spherically symmetric particle-gas potential.

The $\ord{y}$ contribution is vanishing owing to the orthogonality of the stationary and fluctuation modes, \eq{orth}. A straightforward calculation leads to the $\ord{y^2}$ part of the influence functional,
\bea\label{continfly}
S^{(2)}_{infl}&=&\int\frac{d\omega}{2\pi}\biggl\{\tilde{\v{y}}^d_{-\omega}[K^n_\omega(\v{0})-K^n_0(\v{0})]\tilde{\v{y}}_\omega-\tilde{\v{y}}_{-\omega}K^f_\omega(\v{x}^d)\tilde{\v{y}}^d_\omega\nn
&&+\tilde{\v{y}}_{-\omega}i[K^i_\omega(\v{0})-K^i_\omega(\v{x}^d)-K^i_0(\v{0})+K^i_0(\v{x}^d)]\tilde{\v{y}}_\omega\nn
&&+\frac14\tilde{\v{y}}^d_{-\omega}i[K^i_\omega(\v{0})+K^i_\omega(\v{x}^d)-K^i_0(\v{0})+K^i_0(\v{x}^d)]\tilde{\v{y}}^d_\omega\biggr\}+\ord{\dt^2},
\eea
with a vanishing $\ord{\dt}$ piece where the matrices
\be
\hat K_{jk\omega}(\v{x})=\int\frac{d\omega}{2\pi}e^{i\omega t}\nabla_j\nabla_k\hat\Gamma(t,\v{x}),
\ee
have been introduced. One expands $K_\omega$ in $i\omega$ at this point and retains the $\ord{\omega^2}$ terms,
\bea\label{continflyexp}
S^{(2)}_{infl}&=&\int_\omega\biggl[-\frac{\omega^2}2\tilde{\v{y}}^d_{-\omega}\partial^2_{i\omega}K^n_0(\v{0})\tilde{\v{y}}_\omega-\tilde{\v{y}}_{-\omega}i\omega\partial_{i\omega}K^f_0(\v{x}^d)\tilde{\v{y}}^d_\omega+i\frac{\omega^2}2\tilde{\v{y}}_{-\omega}[\partial^2_{i\omega}K^i_0(\v{x}^d)-\partial^2_{i\omega}K^i_0(\v{0})]\tilde{\v{y}}_\omega\nn
&&+\frac14\tilde{\v{y}}^d_{-\omega}i\left(2K^i_0(\v{x}^d)-\frac{\omega^2}2\partial^2_{i\omega}K^i_0(\v{0})-\frac{\omega^2}2\partial^2_{i\omega}K^i_0(\v{x}^d)\right)\tilde{\v{y}}^d_\omega\biggr]+\ord{\dt^2}.
\eea
The influence functional  of the lattice Lagrangian, \eq{dinfl}, assumes the form
\bea\label{dinflex}
S^{latt}_{infl}&=&\int_\omega\biggl[\omega^2\tilde{\v{y}}_{-\omega}\Delta m\tilde{\v{y}}^d_\omega-i\omega(1-i\xi\omega\dt)\tilde{\v{y}}_{-\omega}(k\tilde{\v{y}}^d_\omega+\tilde y^d_{j\omega}\nabla_jk\v{x}^d)\nn
&&+i\frac{\omega^2}2(\tilde{\v{y}}_{-\omega}r\tilde{\v{y}}_\omega+\tilde{\v{y}}^d_{-\omega}q\tilde{\v{y}}^d_\omega)\biggr]+\ord{\dt^2},
\eea
where the $\xi$-dependence, arising from the expansion of $k(\v{x}^d)$, survives the removal of the cutoff, $\dt\to0$, due to the scaling law  $\omega^2y^2_\omega\sim\dt^{-1}$. The matching of the two influence functionals yields the parameters,
\bea\label{matching}
\Delta m_{ij}&=&-\hf\nabla_i\nabla_j\partial^2_{i\omega}\Gamma^n_0(\v{0}),\nn
k_{ij}+\nabla_jk_{i\ell}x^d_\ell&=&\nabla_i\nabla_j\partial_{i\omega}\Gamma^f_0(\v{x}^d),\nn
r_{ij}(\v{x}^d)&=&\nabla_i\nabla_j\partial^2_{i\omega}\Gamma^i_0(\v{x}^d)-\nabla_i\nabla_j\partial^2_{i\omega}\Gamma^i_0(\v{0}),\nn
q_{ij}(\v{x}^d)&=&-\frac14[\nabla_i\nabla_j\partial^2_{i\omega}\Gamma^i_0(\v{0})+\nabla_i\nabla_j\partial^2_{i\omega}\Gamma^i_0(\v{x}^d)],\nn
U_d(\v{x}^d)&=&\Gamma^i_0(\v{x}^d)-\Gamma^i_0(\v{0}),
\eea
where
\be
\Gamma_\omega(\v{x})=\int dte^{i\omega t}\Gamma(t,\v{x}^d).
\ee
In the case of rotational invariance it is advantageous to introduce the functions $\hat\gamma(\v{x}^{d2})=\hat\Gamma(\v{x}^d)$ yielding
\bea
\Delta m&=&-\openone\gamma^n_1-2\v{x}^d\otimes\v{x}^d\gamma^n_2,\nn
k&=&2\gamma^f(\v{x}^{d2})\nn
r&=&2\openone\gamma^i_{1-}(\v{x}^{d2})+4\v{x}^d\otimes\v{x}^d\gamma^i_{2-}(\v{x}^{d2})\nn
q&=&-\hf\openone\gamma^i_{1+}(\v{x}^{d2})-\v{x}^d\otimes\v{x}^d\gamma^i_{2+}(\v{x}^{d2})
\eea
in terms of the coefficient functions,
\bea
\gamma^n_j&=&\partial^j_{\v{x}^{d2}}\partial^2_{i\omega}\gamma^n_0(0),\nn
\gamma^f(\v{x}^{d2})&=&\partial_{\v{x}^{d2}}\partial_{i\omega}\gamma^f_0(\v{x}^{d2}),\nn
\gamma^i_{j\pm}(\v{x}^{d2})&=&\partial^j_{\v{x}^{d2}}\partial^2_{i\omega}\gamma^i_0(\v{x}^{d2})\pm\partial^j_{\v{x}^{d2}}\partial^2_{i\omega}\gamma^i_0(0).
\eea
The dissipation remains isotropic but the decoherence is different in transverse and longitudinal directions, defined by the help of the off-diagonality, $\v{x}^d$.

The matching, \eq{matching}, assures that the perturbative predictions of the full description, based on the action \eq{gasact} and the effective dynamics, defined by the influence Lagrangian, \eq{dinfl}, are identical. Note that the absence of $\ord\dt$ terms in \eq{continflyexp} requires $\xi=0$, which is the midpoint prescription in the discretized Lagrangian. Furthermore, $r=0$ for $\v{x}^d=0$, thus the velocity alone, without the separation of the trajectories can not induce decoherence.

\section{Master equation}\label{masters}
The physical content of the effective Lagrangian is sometime easier to see by means of the corresponding master equation, which is the equation of motion for the density matrix. We present the derivation of this equation for one-dimensional motion by ignoring the difference between the transverse and longitudinal directions in the effective parameters \eq{matching}. The equation of motion for the density matrix can easily be found by calculating the change of the density matrix during a single time step, $t\to t+\dt$,
\be\label{inftst}
\rho(\hx,t+\Delta t)=\frac{m_B}{2\pi\dt\hbar}\int d\hy e^{\ih\dt L_{\dt}(\hx,\hx-\hy)}\rho(\hx-\hy,t),
\ee
in the limit $\dt\to0$ where
\be
L_{\dt}(\hx,\hx-\hy)=\frac{m}{\dt^2}yy^d-U(x^+)-U(x^-)+L_{infl}(\hx,\hx-\hy),
\ee
the last term being given by the one dimensional analog of the influence Lagrangian \eq{dinfl}. We make a further simplification by ignoring the $x^d$ dependence of the effective parameters $\Delta m$, $q$ and $r$.

To identify the relevant terms in the integrand as $\dt\to0$ we inspect the Gaussian integral,
\be\label{toygf}
Z=\int d\hy e^{\frac1{2\hbar\dt}(2imyy^d-qy^{d2}-ry^2)}
\ee
which contains the $\ord{\hy^2/dt}$ part of the effective Lagrangian. The dependence of the density matrix on $\hy$ in the integrand, namely the influence of the initial an final conditions on the path integral plays and important role in \eq{inftst} but it is neglected in \eq{toygf} because it shifts the expectation value of $\hy$ without modifying the fluctuations. By integrating out one of the coordinates,
\bea\label{toygfk}
Z&=&\sqrt{\frac{2\pi\hbar\dt}q}\int d^3ye^{-\frac1{2\hbar\dt}(\frac{m^2}q+r)y^2},\nn
&=&\sqrt{\frac{2\pi\hbar\dt}r}\int d^3y^de^{-\frac1{2\hbar\dt}(\frac{m^2}r+q)y^{d2}},
\eea
one finds a Gaussian distribution for $y$ and $y^d$ and thus the integrals \eq{toygf}-\eq{toygfk} yield
\be\label{jumps}
y^2\sim\frac{q\hbar\dt}{m^2+qr},~~~
y^{d2}\sim\frac{r\hbar\dt}{m^2+qr},~~~
yy^d\sim\frac{m\hbar\dt}{m^2+qr}.
\ee
The Heisenberg canonical commutation relation makes the trajectories $x^\pm(t)$ of a free, isolated particle, $q=r=0$, a nowhere differentiable fractal of Hausdorff dimension two, $(y^\pm)^2=\ord{\dt}$; however, the jumps, $y$ and $y^d$, are correlated, $y^d\sim0$ or $y\sim0$. In fact, if one ignores $x$ or $x^d$ then the time dependence of the other coordinate is driven by the initial or final conditions with negligible local fluctuations. We have $q\ne0$ and/or $r\ne0$ in an open system which make the trajectories $x(t)$ and $x^d(t)$ fractal.

After the reinsertion of the remaining terms of the effective Lagrangian and the density matrix into the integral we expand the right-hand side of eq. \eq{inftst} up to $\ord{\dt}$,
\be\label{expmast}
\rho(\hx,t+\Delta t)=\int d\hy e^{\frac{i}{2\hbar}\hy\hat A\hy}C(\hy)\left[1-\hy\hat\nabla+\hf(\hy\hat\nabla)^2\right]\rho(\hx,t),
\ee
where 
\be
\hat A=\frac1{\dt}\begin{pmatrix}r&-im\cr-im&q\end{pmatrix},
\ee
and $C(\hy)$ is a linear polynomial. Straightforward steps lead to the master equation,
\bea\label{masterg}
\partial_t\rho(\hx,t)&=&\biggl[i\frac\hbar{m_{eff}}\partial\partial_d-\ih U(x^+)+\ih U(x^-)\nn
&&-\frac1\hbar U_{deff}(x^d)+ifx^d\partial+if_dx^d\partial_d+g\partial^2+g_d\partial^2_d\biggr]\rho(\hx,t)
\eea
where $m_{eff}=m+rq/m$ denotes the $x^d$-dependent effective mass, the coefficients are given by $f=kqm/m_{eff}$, $f_d=km/m_{eff}$, $g=\hbar mq/2m_{eff}$, $g_g=\hbar mr/2m_{eff}$ and 
\be
U_{deff}(x^d)=U_d(x^d)-\frac{k^2q}{2(m^2+qr)}x^{d2}
\ee
is an effective decoherence potential. If one retains the $x^d$-dependence of the parameters of the influence Lagrangian one finds as $x^d$-dependent effective mass and the coefficients $f$, $f_d$, $g$ and $g_d$ become sixth order polynomials in $x^d$. 

The second line on the right hand side of eq. \eq{masterg} generate the diffusive part of the effective dynamics. The friction term together with the $\ord{\dot x^2}$ and $\ord{\dot x^{d2}}$ pieces of the influence Lagrangian generate tree-level dynamics and the midpoint prescription in the effective mass produces  one- and the two-loop level contributions. The translation invariance protects against the emergence of $x$ in $\cal L$. The first term on the right-hand side of eq. \eq{masterg}, the kinetic energy, leads to the spread of the wave packet, modulated by the $x^d$-dependence of the effective mass. The decoherence potential, acting on the density matrix picks up mixed effects of the dissipative force and the two-loop level midpoint prescription contributions and governs the stationary decoherence, cf. \eq{zehsme}. The coefficient functions, $f$ and $f_d$, describe a drift of the physical coordinate, $x$, and the quantum fluctuations, $x^d$, respectively. The $\ord{\dot x^{d2}}$ decoherence term of the Lagrangian makes the trajectory $x(t)$ fractal and the emerging operator, $\partial_x^2$, generates diffusion in $x$. The $\ord{\dot x^2}$ part of the Lagrangian is not related to decoherence, it suppresses the momentum, makes $x^d(t)$ fractal and induces diffusion in $x^d$, i.e., recoheres the coordinate. 

The master equation is more restricted for harmonic systems where $\mr{Im}\Gamma^{++}=\mr{Im}\Gamma^{+-}$, cf. eq. \eq{blockg}. It is easy to see that the $\ord{\dot x^2}$ term in the Lagrangian of a harmonic model is proportional to $\mr{Im}(\hat\Gamma^{-1})^{++}-\mr{Im}(\hat\Gamma^{-1})^{+-}$ and thus is vanishing for $\mr{Im}\Gamma^{++}=\mr{Im}\Gamma^{+-}$. Thus the trajectory $x^d(t)$ of the harmonic models is differentiable and there is neither diffusion in $x^d$ nor recoherence. The condition of preserving the positivity of the density matrix \cite{lindblad} requires $g_d=0$ in quadratic models \cite{sandulescuk}.

\end{document}